# Study of Physical Characteristics of the New Half-Heusler Alloy BaHgSn by DFT Analysis

**A. Jabar [1,2], S. Benyoussef [3] and L. Bahmad [3,\*]**

[1] LPMAT, Faculty of Sciences Aïn Chock, Hassan II University of Casablanca, B.P. 5366 Casablanca, Morocco

[2] LPHE-MS, Science Faculty, Mohammed V University in Rabat, Morocco

[3] Laboratory of Condensed Matter and Interdisciplinary Sciences (LaMCScI), Faculty of Sciences, Mohammed V University, Av. Ibn Batouta, B. P. 1014 Rabat, Morocco

*Corresponding author: l.bahmad@um5r.ac.ma (L.B.)

**Abstract**

To investigate the physical characteristics of the half-Heusler BaHgSn molecule, we used theoretical calculations within the Density Functional Theory (DFT) framework utilizing the LSDA+mBJ technique in this study. Using the optimal lattice parameters, we discover that half-Heusler BaHgSn exhibits a Dirac semimetal behavior with a band gap of 0.1 eV. Thomas Charpin's numerical first-principles calculation approach was applied to determine the elastic constants of hexagonal BaHgSn alloys. The material's optical characteristics verified its prospective use in infrared-visible devices. According to a thermo-electric properties analysis, at $20\times10^{18}$ $\Omega^{-1}.m^{-1}.s^{-1}$, the electrical conductivity reaches its maximum after increasing gradually up to 500 K. Compared to other compounds, these results indicate that BaHgSn has potential for use in opto-electronic and thermo-electric devices.

1. Introduction

Prolonged efforts have been drawn to Heusler-based compounds because of their intriguing and adaptable characteristics. Heusler compounds are composed of half-Heusler and full-Heusler semiconductors/metals, denoted by the compositions XYZ and XY$_2$Z, respectively (where Z is the major group element acting as the anion and X and Y are metallic elements with varying charges). Because of their enormous potential in diluted magnetic semiconductors, solar cells, spintronics, thermoelectrics, topological insulators and other applications, semiconductor Heusler compounds are particularly appealing [1], [2] and [3].

Half-Heuslers have been synthesized using several methods: molecular beam epitaxy is used to grow CoTiSb [4], arc-melting, ball-milling, and spark plasma sintering are used to synthesize TiNiSn [5], mechanical alloying with 15 hours of milling is used to create a nanostructured half-Heusler NbFeSb alloy [6], and arc-melting combined with high-pressure, high-temperature and high-pressure method is used to synthesize TmNiSb [7].

The bulk of the Half-Heusler compounds, which have the chemical formula XYZ, have been found to have either orthorhombic or cubic F-43m half-Heusler type structure (PtXSn (X = Zr, Hf) [8]) and hexagonal (LiAlSi [9] and TmNiSb [7]). Both cubic and hexagonal phases are present in some XYZ compounds [10].

RhTiP has been discovered to be an indirect-bandgap conductor with $E_g$ = 1.027 eV within TB-mBJ in recent theoretical work [11]. Additionally, semiconducting behavior shows a direct band gap for LiSrP and an indirect band gap for LiSrAs [12]. The PtZrSn and PtHfSn compounds exhibit an indirect bandgap semiconducting nature, with gaps of 1.23 eV and 0.94 eV, respectively, in the PtXSn compounds [8].

LiSrX (X = As, P) exhibits substantial absorption and low reflection in the visible and low ultraviolet ranges, according to the obtained optical results in Ref. [12]. The band gap for the NbFeSb compounds is 0.37 eV, which may be found using a Tauc plot to analyze UV-Vis absorbance spectra [6]. In Ref. [13], the reflectance of BaAgP remains practically constant from visible to ultraviolet range (1.65–12 eV). TiRu$_{1.8}$Sb exhibits a peculiar glass-like thermal transport behavior and has the lowest lattice thermal conductivity (~1.65 Wm$^{-1}$K$^{-1}$ at 340 K) among the half-Heusler phases that have been reported [1]. At room temperature, RhTiP has a maximum Seebeck coefficient of 1380 µV/K and an electronic figure of merit of 0.25 eV [11]. The Seebeck coefficient for n-type doped was found to be approximately 3116.46 µV/K for LiSrP and 3055 µV/K for LiSrAs. At 300K, the figure of merit achieves a value of 0.98 [12].

A low magnitude of lattice thermal conductivity ($\kappa_L$) is observed at room temperature in Ref. [8] for PtZrSn (16.96 W/mK) and PtHfSn (10.04 W/mK) compounds. The figure of merit for PtHfSn is 0.57, while that of PtZrSn is 0.24. Additionally, TiNiSn's thermoelectric characteristics were optimized by adding Cu, and at 773 K, a maximum figure of merit of 0.6 was attained [5]. Electronic and optical properties were determined using the ab initio method in the earliest theoretical research [14,20]. Conversely, the thermoelectric properties matching the experimental results were demonstrated using the semi-classical Boltzmann transport theory in the BoltzTraP code [20,25].

To the best of our knowledge, the physical properties of the BaHgSn compound have not been the subject of any theoretical or experimental investigation. To close the knowledge gaps about the physical characteristics of such materials, this study was written. For the novice, this paper will offer new paths. The structure of this paper is as follows: We describe the details of the computational method in section 2. We examine the results of the physical properties of newly studied materials in section 3. The conclusions are covered in section 4.

## 2. DFT approach

Using the full-potential linearized augmented plane wave (FPLAPW) technique with a dual basis set, the Wien2k program was used to analyze the system's electronic and optical in detail [26]. For the exchange-correlation interactions, the LSDA+mBJ local density approximation functional was used in the computations [27]. Despite being the most effective theory for analyzing experimental data, LSDA and GGA have a flaw in their ability to anticipate the features of excited states. So, we have employed the mBJ potential to obtain a more accurate representation of the band gaps. The size of the basis sets was determined by the parameter $R_{MT} \times K_{max}$, where $R_{MT}$ signifies the smallest muffin tin radius within the unit cell, and $K_{max}$ corresponds to the magnitude of the largest K vector in reciprocal space. Our calculations expanded the basic functions up to $R_{MT} \times K_{max} = 9$, while Brillouin zone integration utilized a modified tetrahedron method with a total of 1000 k-points. For the computation of thermoelectric coefficients, a combined approach involving first-principles band structure calculations and Boltzmann transport theory within the rigid band approximation (RBA) and constant scattering time approximation (CSTA) was employed, utilizing the BoltzTrap code [28]. The crystal structure of Half-Heusler Hexagonal BaHgSn (P63/mmc, space group 194) is shown in Fig. 1 [29]. At the spatial origin is the Ba ion. The trigonal layers of Ba ions are intercalated between the Hg and Sn ions to generate AA' stacking honeycomb layers. Table 1 lists the principal crystallographic positions.

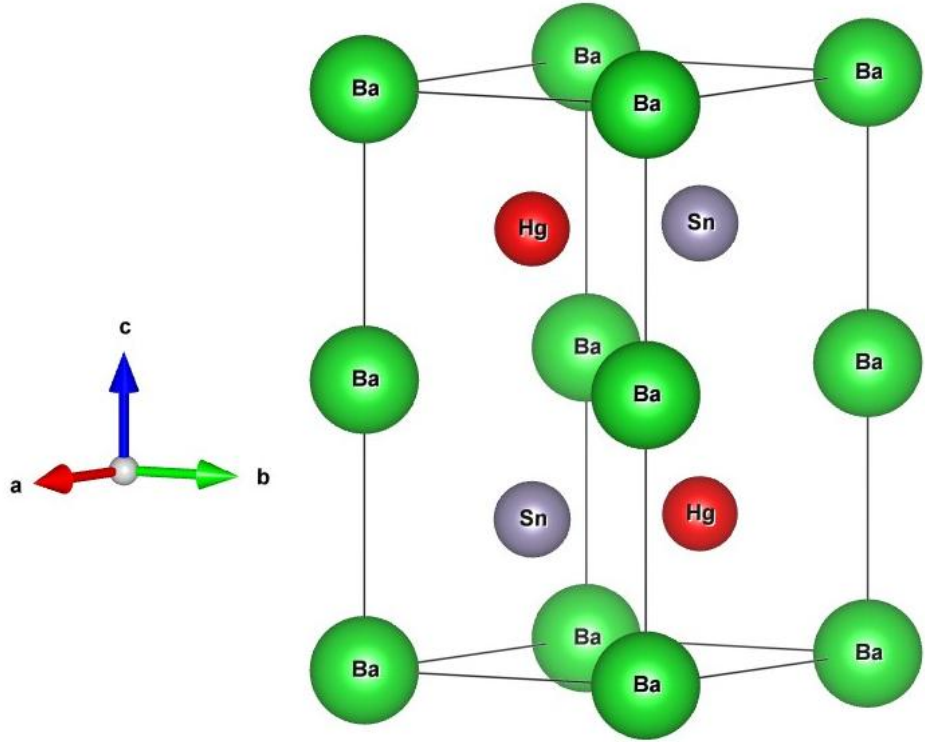

*Fig. 1*: *The crystal structure of BaHgSn.*

*Table 1:* *Crystallographic position for BaHgSn Phase in the trigonal System*

|    | x       | y       | Z       |
|----|---------|---------|---------|
| Ba | 0.33333 | 0.66667 | 0.25000 |
| Hg | 0.00000 | 0.00000 | 0.00000 |
| Sn | 0.33333 | 0.66667 | 0.75000 |

### 3. Results and discussions

#### 3.1. Total energy and lattice parameters

The bulk modulus, lattice constants, and static pressure transitions are determined via structural optimization before studying the electronic transport properties of half-Heusler BaHgSn. The empirical Birch-Murnaghan equation of state is used to carry out the optimization [30]. The relationship between a substance's volume and energy is explained by the Murnaghan formula. It is frequently used to assess the energy volume curve that results from calculations to find the bulk modulus and equilibrium lattice parameters of a material. The Murnaghan equation is defined as:

$$E(V) = E_0(V) + \left[\frac{B_0 V}{B'_0(B'_0 - 1)}\right] * \left[B_0\left(1 - \frac{V_0}{V}\right) + \left(\frac{V_0}{V}\right)^{B'_0} - 1\right]$$

Where $V_0$ is the volume of the unit cell in its ground state, $B_0$ is the bulk modulus, and $B'_0$ is the pressure derivative of that modulus coupled to the unit cell's volume.

The optimized parameters for the studied compound are obtained, including the lattice constants ($a_0$, $b_0$ and $c_0$), volume ($V_0$), bulk modulus ($B_0$), its first pressure derivative (B'), minimum total energy ($E_0$) and the gap energy ($E_g$). The values of these parameters are summarized in Table 2. The computed lattice parameters match the experimental values quite well [31]. Additionally, the energy as a function of the volume curves is depicted in Fig. 2.

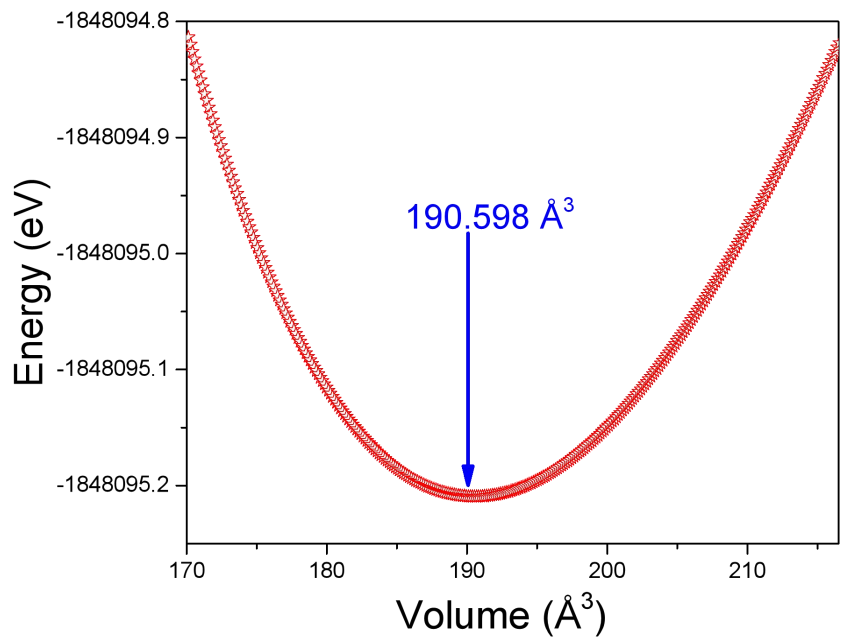

***Fig. 2****: The energy vs. the volume of the BaHgSn compound.*

**Table 2:** *Calculated equilibrium lattice parameters, such as lattice constants $a_0$ and $c_0$ (in Å), volume $V_0$ (in Å³), bulk modulus $B_0$ (in GPa), its first pressure derivative B', the minimum total energy $E_0$ (in eV) and the gap energy $E_g$ (in eV) for the half-Heusler BaHgSn compound by employing the LSDA+mBJ approximation.*

| $a_0$(Å) | $c_0$(Å) | $V_0$(Å³) | B(GPa) | B' | $E_0$(eV) | $E_g$(eV) |
|---|---|---|---|---|---|---|
| 4.889 | 9.613 | 190.598 | 45.687 | 4.997 | -1848095.210 | 0.1 |

### 3.2. Elastic constants

To have a more comprehensive theoretical understanding of the properties of materials that are dictated by the density of states of phonons and the processes of electron–phonon interaction, one must be familiar with elastic constants. Moreover, they have a close relationship with micro-hardness and mechanical characteristics. Practical applications can assume a linear relationship between stress and strain because the stress-strain curve in many solids shows negligible nonlinearity.

To anticipate the elastic constants of half-Heusler BaHgSn alloys, we employed a numerical first-principles calculation method developed by Thomas Charpin, which is integrated into the Wien2k package, as detailed in Ref. [26]. This approach entails applying minor strains to the unstrained lattice to compute the elastic constants. In the case of Hexagonal symmetry, six independent elastic constants, denoted as $C_{11}$, $C_{12}$, $C_{13}$, $C_{33}$, $C_{44}$, and $C_{66}$, exist. To ascertain these constants, six distinct strains are necessary. It is noteworthy that the $C_{ij}$ constants for all systems adhere to the generalized criteria [32] for mechanically stable crystals, specifically $C_{44} > 0$, $C_{11} > |C_{12}|$, and $(C_{11} + 2C_{12}) C_{33} > 2 C_{13}^2$. The calculated values of the $C_{ij}$ constants are presented in Table 3, indicating mechanical stability. The derived elastic moduli values for the Hexagonal BaHgSn compound serve as a foundational reference for upcoming projects involving this material.

**Table 3:** The calculated elastic constants $C_{ij}$ in the unit of (GPa) for BaHgSn.

| $C_{11}$ | $C_{12}$ | $C_{13}$ | $C_{33}$ | $C_{44}$ | $C_{66}$ |
|---|---|---|---|---|---|
| 180.9 | -29.5 | 30.6 | 42.7 | 32.2 | 105.2 |

### 3.3. Electronic properties

Using relaxed crystal structures, we computed the electrical structures. We calculated the partial density of states (PDOS) and the total density of states (TDOS) for Ba, Hg, and Sn using the LSDA+mBJ approximation. The results are shown in Figures 3(a), 3(b), 3(c), and 3(d), respectively. The hybridization of the Sn-p and Hg-p electrons results in the valence bands. However, for half-Heusler BaHgSn, Ba-d, and Sn-p dominate the conduction bands. The compound is also non-magnetic due to the symmetric spin-up and spin-down contributions within the TDOS. Fig. 3(e) shows the band structure plot of the majority and minority spin channels using LSDA-mBJ as the exchange-correlation potential. In the Brillouin Zone, electronic band structures are depicted along the high symmetry directions along the path Γ–M–K–Γ–A. It is noteworthy that this material exhibits a Dirac semimetal which can host hourglass-like surface states. Also, BaHgSn has an additional band inversion near Γ point. This band inversion is induced by the stronger inter-layer coupling among Hg-Sn honeycomb layers, leading to bulk Dirac nodes along the layer stacking direction Γ-K. Our results demonstrate the studied compound is semimetal with a small band gap of 0.1 eV at the Dirac cone. These outcomes mostly resemble those of earlier research [11], [19].

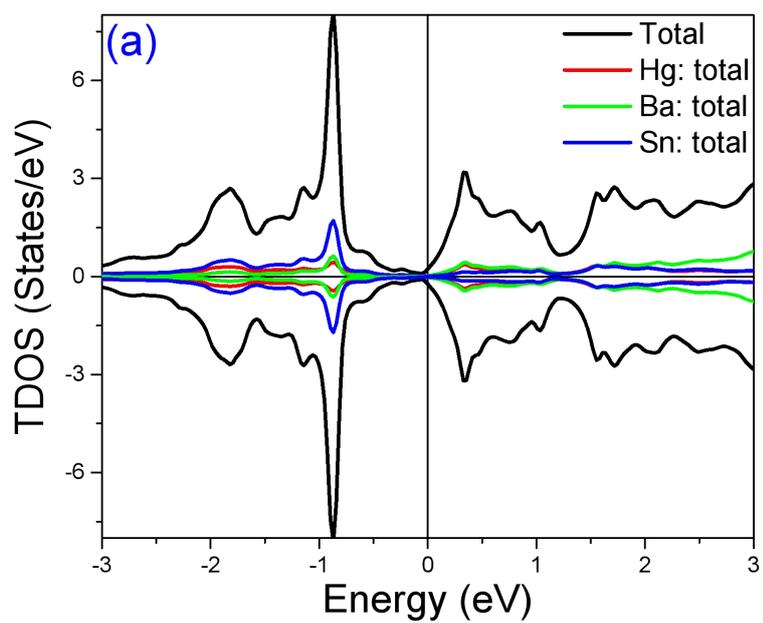

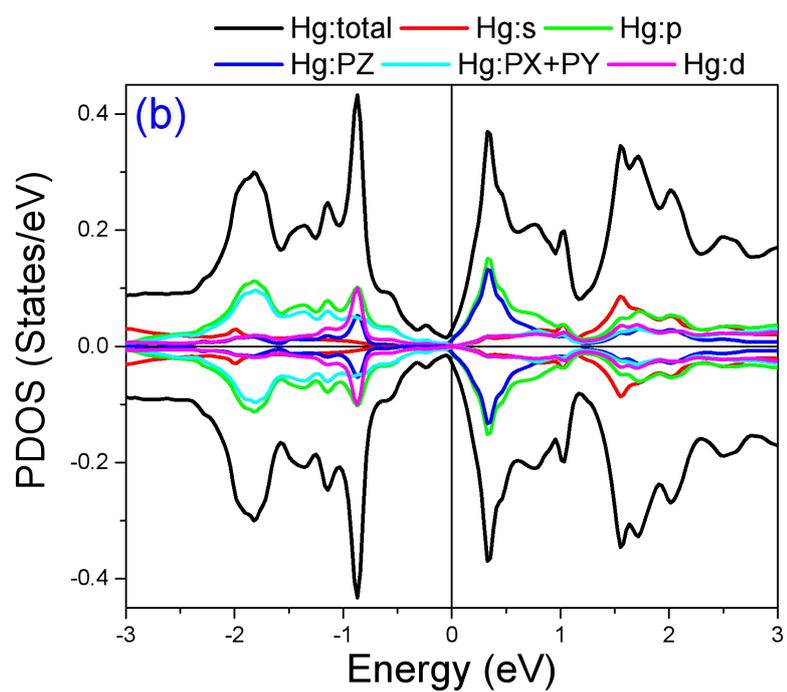

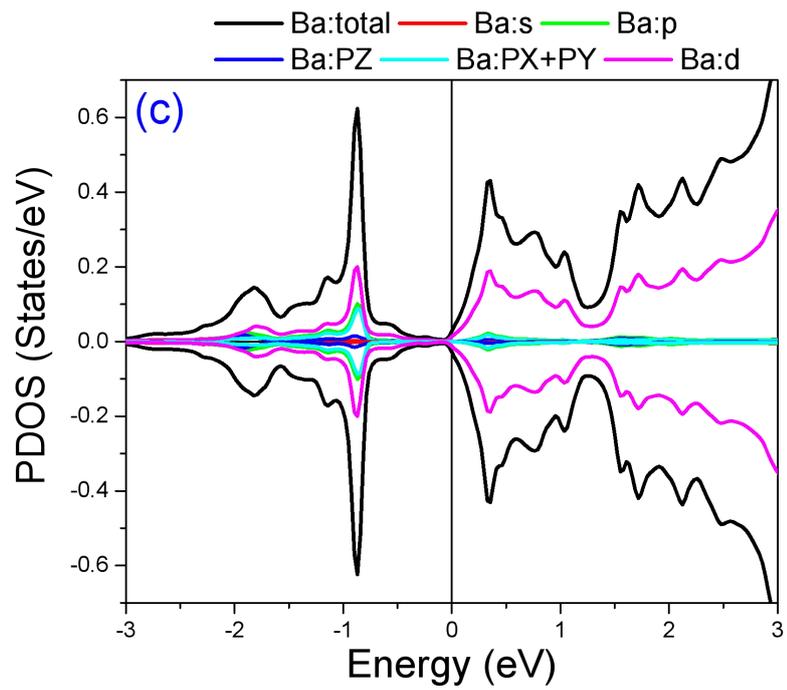
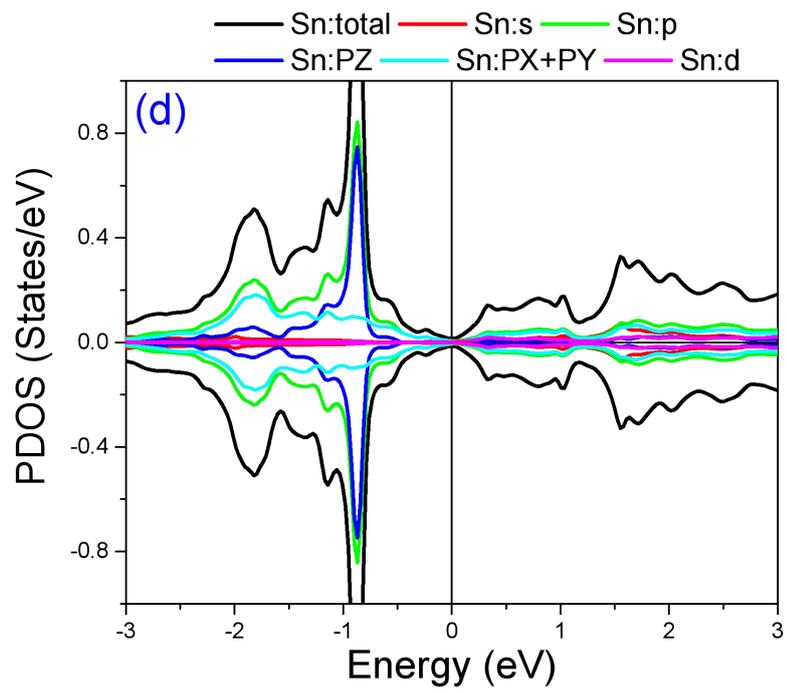

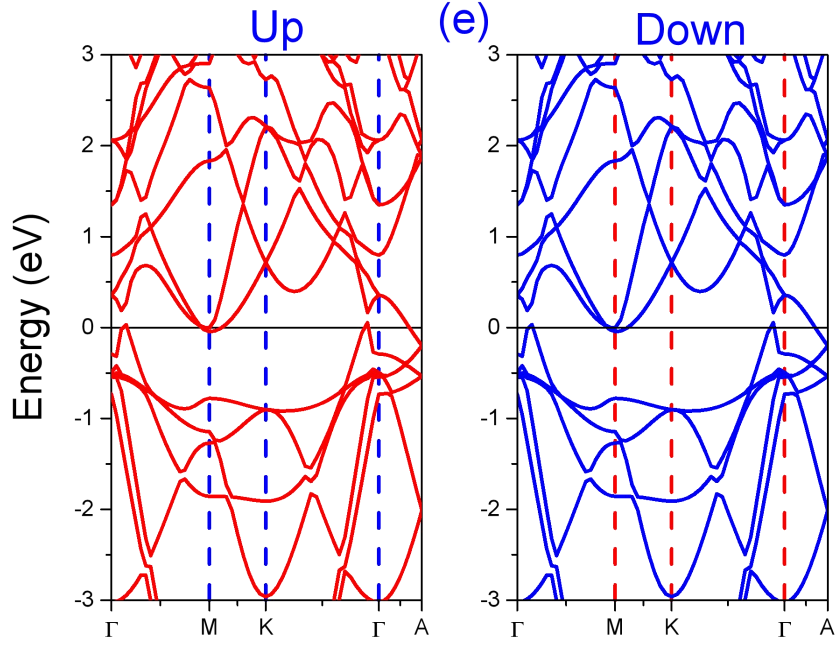

***Fig. 3:*** *The total density of state (TDOS) (a), the partial density of state (PDOS) (b, c and d) and the band structure up and down (e) of BaHgSn using LSDA+mBJ approach.*

### 3.4. Optical properties

In materials research, the study of a compound's optical characteristics has become essential. It's also critical to investigate potential uses in the photovoltaic and optoelectronic device industries. The primary way that an electromagnetic wave interacts with a material's charge carrier is described by its optical characteristics. Its optical properties are analyzed in detail by looking at factors such as the refractive index, absorption coefficient, energy electron loss, real and imaginary components of the dielectric tensor, and optical conductivity. The results of this study are illustrated in Figs. 4 in the xx, yy direction and subsequently along the zz direction of the half-Heusler BaHgSn compound using the LSDA+mBJ method.

The half-Heusler BaHgSn compound's refractive index in the energy range of 0–13,5 eV is shown in Fig. 5(a), where the static refractive index is 3.7 along the zz direction and 4.75 along the xx, yy direction. The curves coincide and increase in value as energy levels rise, reaching their maximums at 6.39 at 0.5 eV along the xx, yy direction and 4.87 at 1.2 eV along the zz direction. Because of the electronic polarization that occurs when photons pass through the former's medium, it is suggested that the photons are more slowed by the greater refractive index computed along the xx, yy direction than that along the zz direction. BaHgSn is suitable for solar applications because of these properties.

The energy that an electron loses when traveling quickly through a substance is described by the energy-loss function. The electron energy loss function of the half-Heusler BaHgSn molecule is shown in Fig. 4(b) along the xx, yy direction and along the zz direction. The curves demonstrate that, for photon energies between 0 and 1 eV, the energy-loss function for both directions is zero. It then begins to increase with increasing photon energies, peaking at 13.2 and 13.3 eV along the xx, yy direction and along the zz direction, respectively. It is noteworthy that the UV spectrum exhibits the largest energy loss. These discrete peaks are the result of inter-band transitions between discrete points of high symmetry.

Understanding a material's electrical nature whether it is metallic, semiconducting, or insulating requires knowledge of its absorption coefficient. It also aids in our comprehension of a material's ideal solar energy conversion efficiency. The absorption coefficient of the half-Heusler BaHgSn compound is shown in Fig. 4(c) along the xx, yy direction and along the zz direction. Optical absorption, as we can see, begins at zero photon energy. The unbound electrons in the conduction band are what cause the optical absorption to begin. In the range of around 1.5 to 8 eV, the absorption coefficient is relatively high in the xx, yy direction, peaking at 6.05 eV. In the spectral band between 4.6 and 7.5 eV, the absorption coefficient is strong along the zz direction, peaking at 7.3 eV.

Fig. 4(d) shows the real and imaginary parts of the dielectric tensor along the xx, yy direction and along the zz direction of the half-Heusler BaHgSn compound. The degree of polarization and the distribution of photons within the material are explained by the dielectric tensor's real portion. The dielectric tensor's real part rises with increasing energy at first, reaching its maximum values along the xx, yy, and zz directions at 0.15 eV and 1.31 eV, respectively. This is followed by a sharp decline to the negative values at 1.62 eV and 5.2 eV energy, respectively, and subsequent fluctuation along the xx, yy, and zz directions. It is discovered that the static dielectric function is the same in both directions and equal to 13.

The maximum value of 16 at 1.4 eV and 35 at 2.1 eV along the zz direction and xx, yy direction, respectively, indicate that the absorption is maximum in the Infrared (IR)-visible (Vis) region. The imaginary part of the dielectric tensor begins decreasing from zero at around 0.1 eV, which is an absorption edge. These curves begin to fluctuate after an abrupt decline. These materials show promise in optoelectronic devices, like electro-optic sensors, due to their absorption in the IR–Vis region.

The optical conductivity of the half-Heusler BaHgSn compound is compared in Fig. 4(e), where the real and imaginary parts of the conductivity are shown along the xx, yy, and zz directions. The conductivity reaches its maximum at 1.53 eV along the xx, yy direction, and along the zz direction, respectively, according to the real component of optical conductivity. As shown by Fig. 4(e), the conductivity along the xx, yy direction is at lower energy than that along the zz direction. This is because the former has a lower bandgap energy than the latter, which makes it easier for electrons to move from the valence band to the conduction band (at lower energy).

The part of optical conductivity that is imaginary and describes how the applied field is screened. In both directions, the imaginary part of the optical conductivity drops from 0 to −3.

They both see a sharp decline after reaching their maximum values at 2.7 and 2.3 in the xx, yy direction and along the zz direction, respectively.

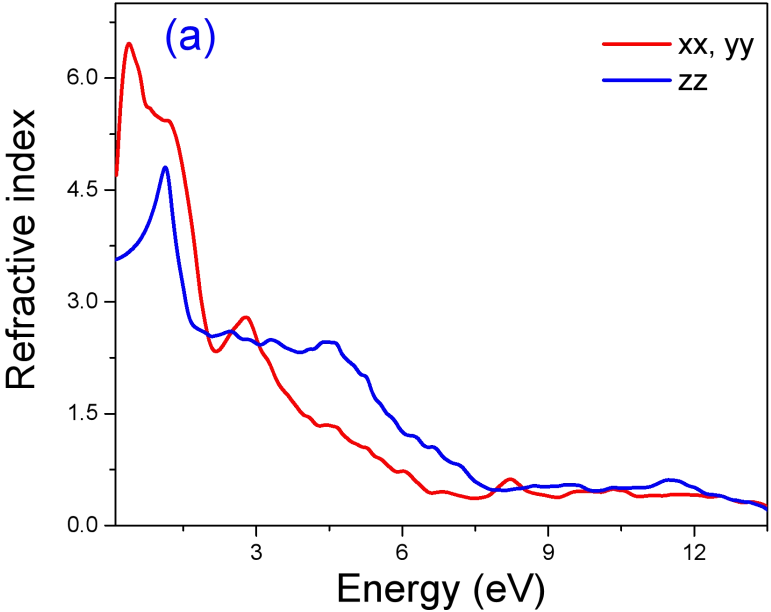

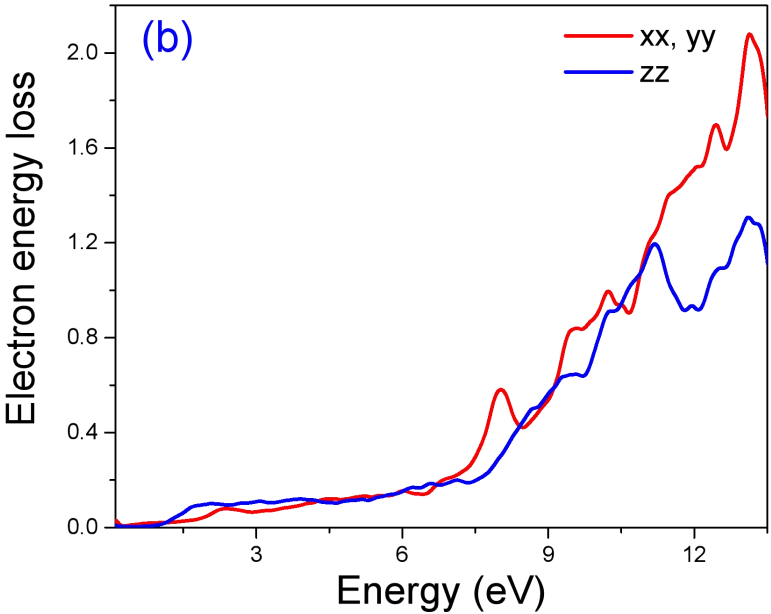

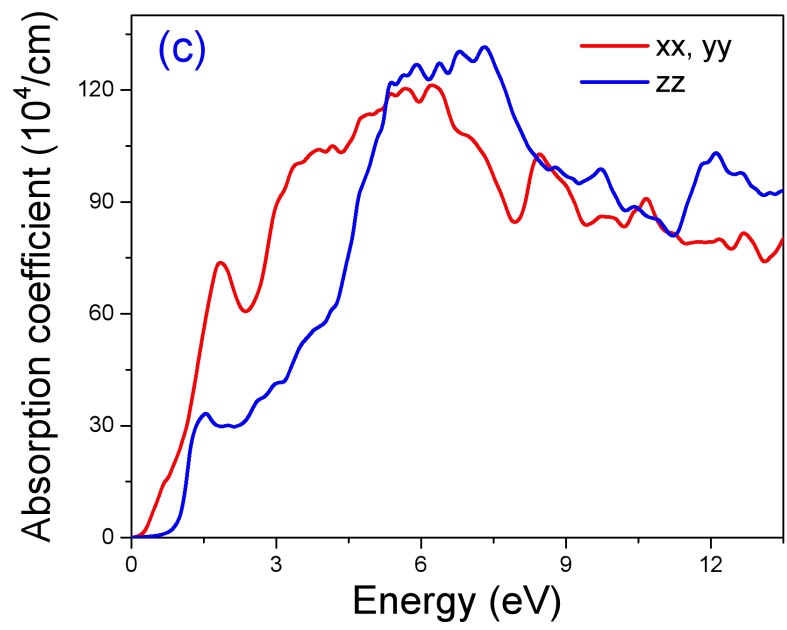

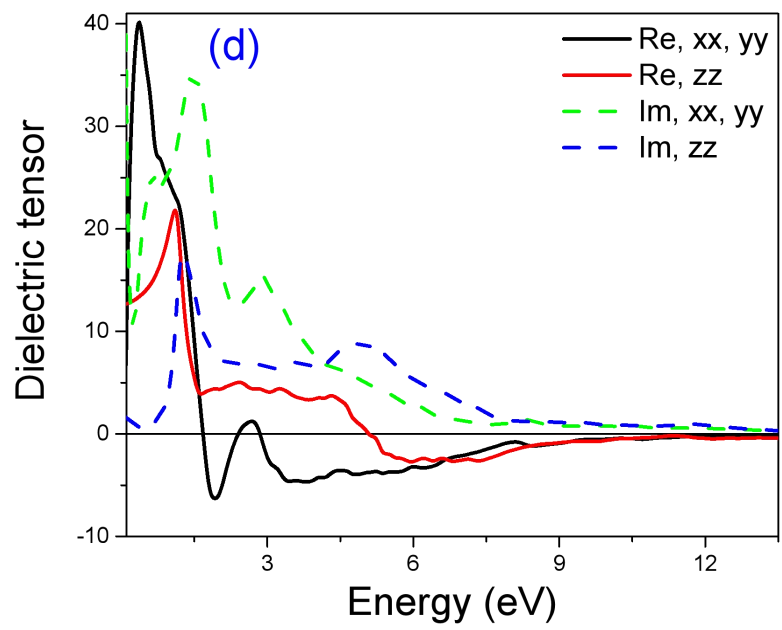

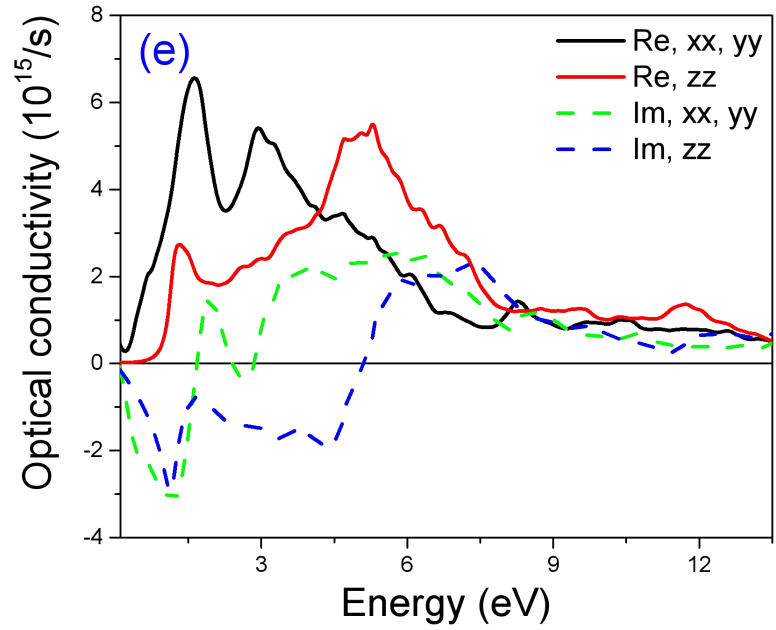

***Fig. 4***: *Optical properties of half-Heusler BaHgSn: the refractive index (a), the electron energy loss (b), the absorption coefficient (c), the dielectric tensor (d) and the optical conductivity (e) by using the LSDA+mBJ approach.*

### 3.5. Thermoelectric properties

Thermoelectric properties pertain to the inherent characteristics of materials that dictate their capacity to convert heat energy into electrical energy, and vice versa, through the phenomenon known as the thermoelectric effect. The transport properties of the half-Heusler BaHgSn compound were delineated by assessing the Seebeck coefficient (S), electrical conductivity ($\sigma/\tau$), and the electronic component of thermal conductivity ($\kappa_e/\tau$). Across the temperature range of 0 to 500 K, these properties were determined utilizing the Boltzmann transport equation for electrons, incorporating a consistent scattering time [28]. The Seebeck coefficient (S) of half-Heusler BaHgSn, depicted in Fig. 5(a), demonstrates a rapid decrease up to 55 K, followed by a gradual increase. Fig. 5(b) illustrates the variation of electrical conductivity with temperature. The electrical conductivity ($\sigma/\tau$) experiences a gradual increase up to 500 K, explained by the appearance of the band structure of half-Heusler BaHgSn. The maximum calculated electrical conductivity of half-Heusler BaHgSn at 500 K is $20\times10^{18}$ $\Omega^{-1}\cdot m^{-1}\cdot s^{-1}$, a value reasonably close to that of TiPdSn [33].

The computation of electronic thermal conductivity ($\kappa_e/\tau$) is derived from the electrical conductivity σ through the Wiedemann-Franz relationship, $\kappa_e = L\sigma T$, where L represents the Lorenz number (approximately $2.44\times10^{-8}$ $W\cdot\Omega\cdot K^{-2}$) [34]. The electronic part of thermal conductivity ($\kappa_e/\tau$) at different temperatures is depicted in Fig. 5(c). It is evident that electronic thermal conductivity also experiences a gradual increase up to 500 K. The maximum calculated electronic thermal conductivity at 300 K is 17.41 W/mK, which closely aligns with the value for TiPdSn (20 W/mK) [33]. Consequently, half-Heusler BaHgSn emerges as a promising candidate for thermoelectric devices. Additionally, we computed electronic specific heat, Pauli magnetic susceptibility, and lattice thermal conductivity using the LSDA+mBJ approach. The electronic heat capacity (C) of BaHgSn in the temperature range from 0 to 500 K is displayed in Fig. 5(d). The curve indicates that in the temperature range of 0 to 10 K, the main contribution to electronic-specific heat comes entirely from the excitation of electrons. As depicted in the figure, with increasing temperature, the specific heat also rises. However, from 10 to 500 K, the contribution to C also arises from the excitation of phonons. Fig. 5(e) presents the Pauli magnetic susceptibility (χ) as a function of temperature for the BaHgSn compound. The Pauli magnetic susceptibility slightly decreases, reaching its lowest value at 20 K, before increasing again. The maximum value of χ is 3.87 $m^3\cdot mol^{-1}$ at 75 K. Notably, the Pauli magnetic susceptibility is consistently low for this non-magnetic material with a closed-shell electronic configuration, where all electrons are paired, resulting in a net zero magnetic moment.

The lattice thermal conductivity $\kappa_L$ is determined using the Debye temperature ($\theta$) and the Grüneisen coefficient ($\gamma$), obtained from computations using the Gibbs2 software programs [35]. The methodology also incorporates an exceptional relationship to calculate the thermal conductivity $\kappa_L$, utilizing the Slack approximation outlined by the equation:

$$\kappa_L = \frac{A\theta_D^3 V^{1/3} M}{\gamma^2 n^{2/3} T} \qquad (6)$$

Where A is a group of physical parameters. According to [35], this can be computed as

$$A = \frac{2.43 \times 10^{-8}}{1 - 0.514/\gamma + 0.228/\gamma^2} \qquad (7)$$

$\theta_D$ is the temperature of Debye, $\gamma$ is the Grüneisen coefficient, V is the volume by atom, T is the temperature, n is the number of atoms in the primitive unit cell, and M is the atomic mass.

As shown in Fig. 5(f), the thermal conductivity of the lattice ($\kappa_L$) for the BaHgSn utilizing the LSDA+mBJ method increases with pressure and decreases exponentially with temperature. The material exhibits an anti-harmonic effect, as indicated by the high value of $\kappa_L$. We can say that the current material may be highly promising for thermoelectric utility because of the tiny lattice thermal conductivity value. They could not be compared in the absence of any experimental data or theoretical conclusions.

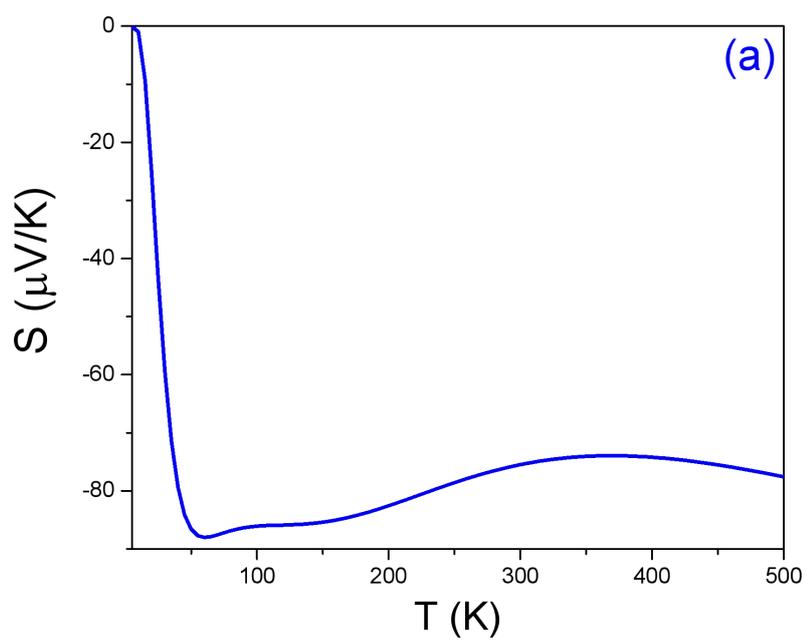

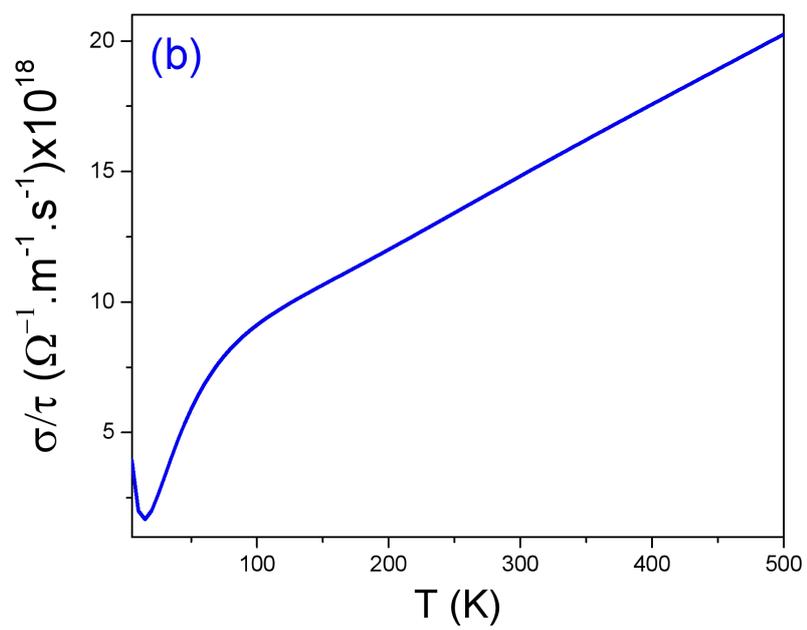

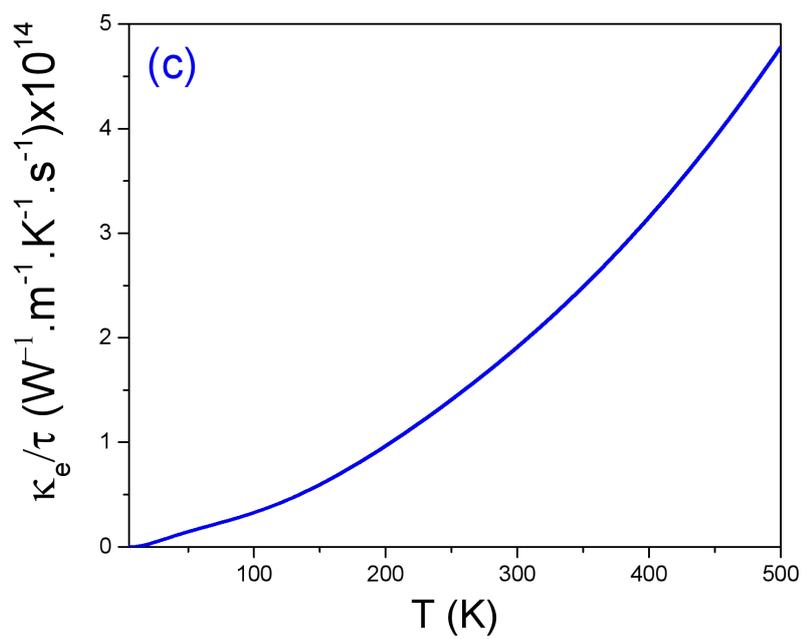
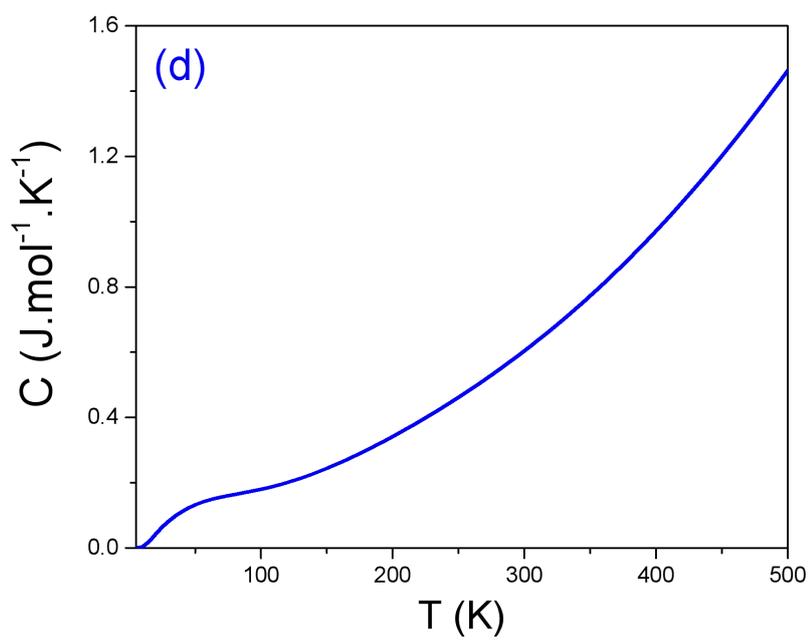

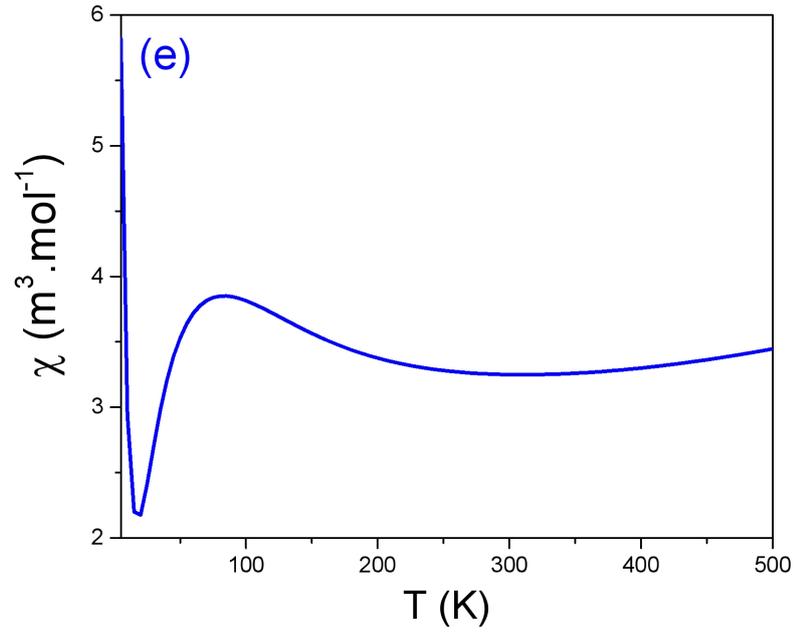

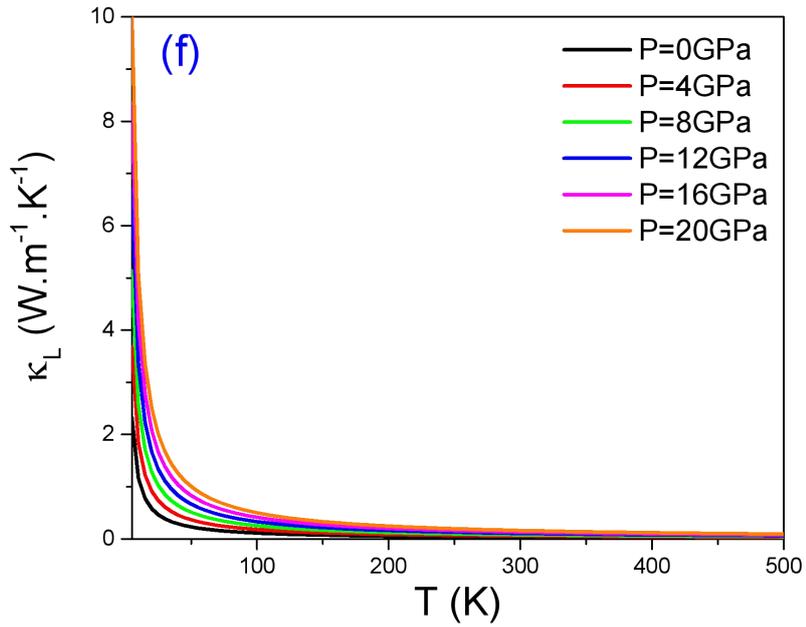

*Fig. 5: Thermoelectric properties of BaHgSn: (a) Seebeck coefficient (S), (b) electrical conductivity (σ/τ), (c) electronic part of the thermal conductivity ($\kappa_e/\tau$), (d) the electronic specific heat (C), (e) the Pauli magnetic susceptibility (χ) and (f) the lattice thermal conductivity ($\kappa_L$) by using the LSDA+mBJ approach.*

## 4. Conclusion

In summary, we used density functional theory (DFT) and semi-classical Boltzmann transport theory incorporated in Wien2k to study the physical characteristics of half-Heusler BaHgSn. The energy-volume curve, fitted with the Birch–Murnaghan equation of state, showed that the materials under study were structurally stable. The material's semimetal character was disclosed by the computed electronic properties, which showed a bandgap energy of 0.1 eV in the Dirac cone for the half-Heusler BaHgSn combination. The bandgap energy estimates were likewise consistent in the calculated total and partial density of states. These materials show promise for optoelectronic applications due to their optical properties, which include dielectric function, optical conductivity, refraction index, energy loss function, and absorption coefficient. These qualities suggest that the material may find use in IR-vis devices. Additionally, the mechanical stability of this material was discovered. Moreover, the n-type of the material under study was identified using a thermoelectric property investigation. At 300 K, the greatest electronic thermal conductivity that has been calculated is 17.41 W/mK. Compared to other compounds, these results indicate that half-Heusler BaHgSn has potential for use in thermoelectric devices. We expect that this work will bring experimental studies to the attention of future investigators.